\documentstyle[sprocl,fleqn,epsfig]{article}

\def\be{\begin{equation}}
\def\ee{\end{equation}}
\def\bea{\begin{eqnarray}}
\def\eea{\end{eqnarray}}
\newcommand{\dcs}{\mbox{$\sigma_{d}$}}
\newcommand{\aem}{\mbox{$\alpha_{em}$}}
\begin{document}

\title{COLOUR DIPOLES and SATURATION}

\author{J. FORSHAW, G. KERLEY and G. SHAW}

\address{ Department of Physics 
        and Astronomy,The University of Manchester,\\
Manchester M13 9PL, U.K.\\E-mail: graham.shaw@man.ac.uk}


\maketitle\abstracts{We employ  values of the colour dipole cross 
section  extracted from electroproduction and photoproduction data 
to show that gluon saturation effects are not required by the current HERA
data but will become important in the THERA energy region. }

\section{The colour dipole model}

Singly dissociative diffractive $\gamma$ p processes are conveniently 
described in the rest frame of the hadron using a picture in which the 
incoming photon dissociates into a hadronic state which subsequently 
interacts with the proton.
%undergoes an initial fluctuation into a $q \bar{q}$ pair
%a long distance - typically of order of the ``coherence length'' $1/Mx$ 
%- from the target proton.  Assuming that the resulting partonic/hadronic 
%state evolves slowly compared to the size of the proton or nuclear target,
%it can be regardes as frozen during the interaction. In this approximation,
%the process will factorize into a probability for the photon to have evolved 
%into a given state $ | \alpha >$, times the amplitude for that state to 
%interact with the target.
In the colour dipole model, ~\cite{bib:dcs_nik,bib:dip_muel_1}
the dominant states  are assumed to 
be $q \bar{q}$ states of given transverse size. Specifically 
\begin{equation}
  \label{eq:photon_wf}
  |\gamma\rangle = \int \mbox{d}z \mbox{d}^{2}r \ \psi (z,r) |z,r\rangle + 
\ldots \; ,
\end{equation}
where $r$ is the transverse size of the pair, $z$ is the fraction of light 
cone energy carried by the quark and $\psi (z,r)$ is the \emph{light cone 
wave function} of the photon. Assuming that these states are  scattering
eigenstates (i.e. that $z, r$ remain unchanged in diffractive scattering) 
one obtains
%then the elastic scattering amplitude for $\gamma^* p \to \gamma^* p$ is 
%specified by Figure 1. This leads via the optical theorem to  
\begin{equation}
  \label{eq:sigma_tot}
   \sigma^{\gamma^{*}p}_{T,L} = \int \mbox{d}z \mbox{d}^{2}r \ |\psi_{\gamma}^
{T,L}(z,r)|^{2} \sigma(s,r,z) \; , 
\end{equation}
for the $\gamma^* p$ total cross-section in deep inelastic scattering,
where  $\sigma(s,r,z)$ is the total cross-section
for scattering dipoles of specified  $(z,r)$  from a proton at 
fixed $ s = W^2$. This ``dipole cross-section'' is a universal quantity for 
singly-dissociative 
diffractive  processes on a proton target, playing a similarly 
fundamental role in, for example, open diffraction, exclusive
 vector meson production and deeply virtual Compton scattering.

%\begin{figure}[htb]
%\begin{center}
%\psfig{file=protonsf.eps,width=7cm,height=3cm}
%\caption{The colour dipole model for $\gamma^* p \to \gamma^* p$.
%\label{fig:fluct}
%}
%\end{center}
%\end{figure}

\begin{figure}[htb]
\begin{center}
\psfig{file=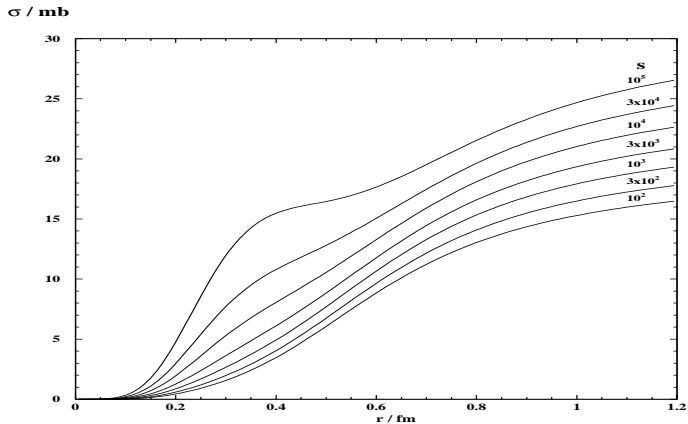, width=10cm,height=6cm}
\caption{The dipole cross-section as a function of s in the HERA range.
\label{fig:dcs}
}
\end{center}
\end{figure}

\begin{figure}
\begin{center}
\psfig{file=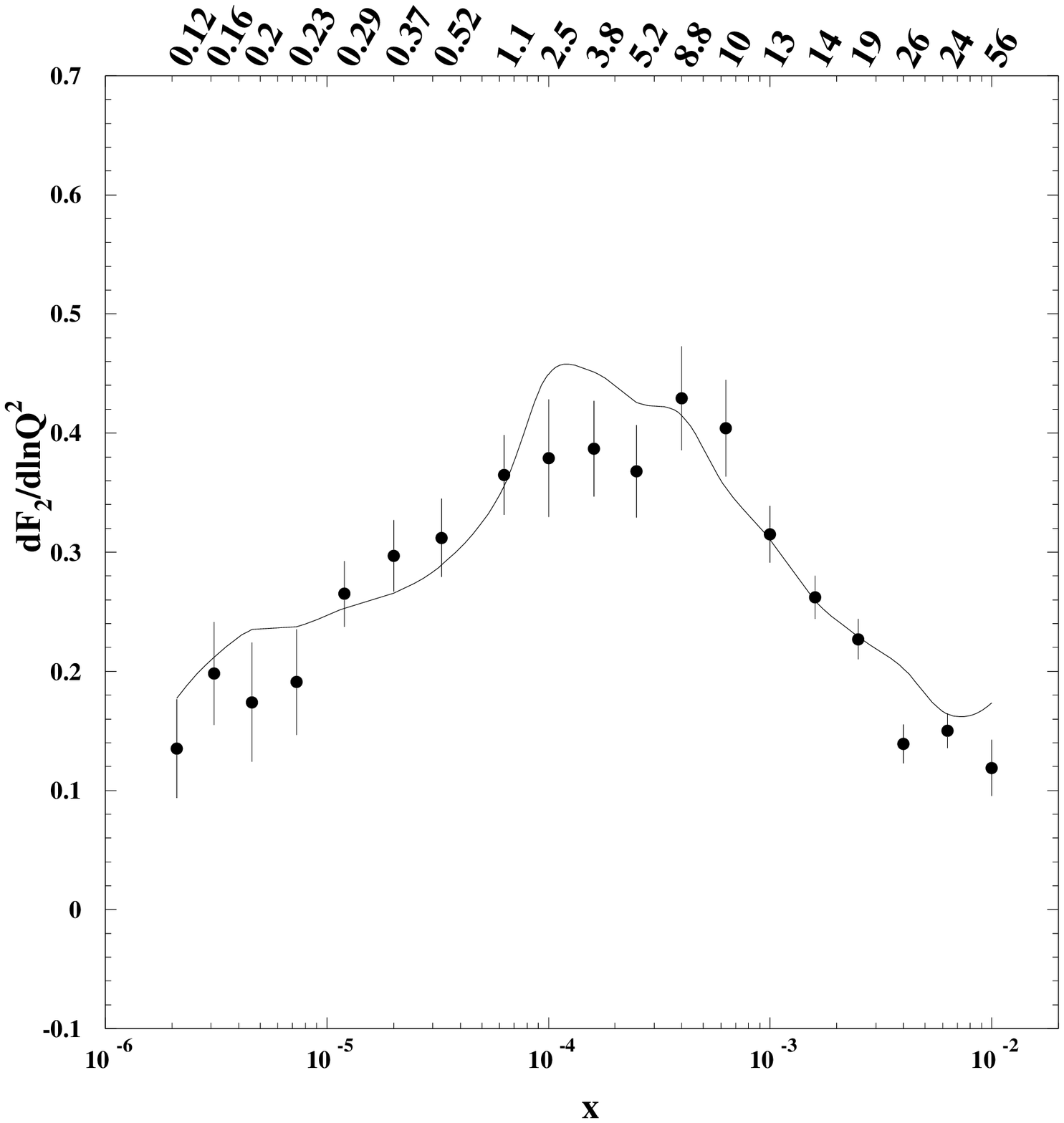, width=10cm,height=6cm}
\caption{The Caldwell plot. Predictions are made at the $Q^2$ values of
the data points and roughly interpolated.
\label{fig:caldwell}
}
\end{center}
\end{figure}

\section{The dipole cross-section}

We have extracted the dipole cross-section
from DIS and real photoabsorption data assuming a form with two 
terms with a  Regge type $s$ dependence and no dependence on $z$:
\begin{equation}
\label{gammatot}
  \sigma(s,r)  =   a_{soft}(r) s^{\lambda_{S}} + a_{hard}(r) s^{\lambda_{H}} 
\; \end{equation}
%\begin{equation}
%\label{gammatot}
%  \sigma(s,r)  =   \sigma_{soft}(s,r) + \sigma_{hard}(s,r) 
%\end{equation}
%where
%\begin{eqnarray*}
%\sigma_{soft}(s,r)  &=& 
%a_{0}^{S}\left(1 - \frac{1}{1 + a_{4}^{S}r^4}\right)(r^{2}s)^{\lambda_{S}}
%\\
%\sigma_{hard}(s,r)  
%&=& 
%\left(a_{2}^{H}r^{2} + a_{6}^{H}r^{6}\right)
%\exp(-\nu_{H}^{2}r) (r^{2}s)^{\lambda_{H}}
%\end{eqnarray*}
where the values $\lambda_S \approx 0.08$, $\lambda_H \approx 0.42$
resulting from the fit are characteristic of the soft and hard pomeron
respectively. The functions $a_{soft}(r)$, $a_{hard}(r)$ are chosen so
that for  small dipoles
the hard term dominates yielding a 
behaviour $\sigma \rightarrow r^2 (r^2 s)^{\lambda_H}$ as $r \rightarrow 0$
in accordance with colour transparency ideas; while for large dipoles
%\footnote{The 
%behaviour predicted by the above equations for  very large 
%dipoles $r >> 1 f$ should not be taken 
%seriously, since their contribution is exponentially by the wavefunctions}
 $r \approx 1$ fm the soft 
term dominates with a hadronlike behaviour $\sigma \approx \sigma_0 
(r^2 s)^{\lambda_S}$. Further details of our approach, including the treatment
 of the wavefunctions, may be found elsewhere ~\cite{bib:our}. The resulting 
dipole cross-section is shown in Figure 1.

\section{Saturation}

For high enough energies, the assumed $s^\lambda \;( \lambda > 0)$ behaviours 
assumed above  must be tamed by unitarity effects, especially for the hard
term with $ \lambda_H \approx 0.4$. At fixed $Q^2$, $x \rightarrow 0$ as
$s \rightarrow \infty$ and the resulting  softening of the corresponding 
$x^{- \lambda_H}$ behaviour is associated with gluon saturation in the
quark-parton language. The fact that we obtain an excellent fit to
the DIS data, means that the current HERA data are
not at sufficiently high $s$ to {\em require} the saturation effects that are
built into some other similar dipole models ~\cite{bib:gbw,bib:mcd}.
We note that our model agrees
with the standard Caldwell plot Figure 2, where the turn over as $x$
decreases occurs because $Q^2$ is also decreasing; no such effect is predicted
in our model if $x$ is decreased at fixed $Q^2$, as confirmed by the
preliminary  ZEUS97 data.

A strong indication of when saturation effects will be needed is given
in Figure 1. As can be seen, the cross-section for small dipoles
is initially small but increases rapidly and  at the top of 
the accessable HERA range
is becoming commensurate with the
slowly increasing ``hadronic'' behaviour af the large dipoles. It is at this
point that saturation effects are expected to become important;  if they 
don't, the cross-section for small dipoles will exceed that for large dipoles 
at higher energies and the dipole cross-section will paradoxically decrease 
with increasing size $r$. Saturation effects are therefore expected to play
an important role just beyond beyond the HERA range, in the planned THERA 
region with $s_{max} \approx 10^6$ GeV.

\end{document}